\documentclass[journal]{IEEEtran}

\ifCLASSINFOpdf
  \usepackage[pdftex]{graphicx}
  \graphicspath{{../pdf/}{../jpeg/}}
  \DeclareGraphicsExtensions{.pdf,.jpeg,.png}
\else
\fi

\usepackage{cite}
\usepackage{graphicx}
\usepackage{amsmath}
\usepackage{algorithmic}
\usepackage{array}
\usepackage{url}
\usepackage{amssymb}
\usepackage{authblk}
\usepackage{stfloats}
\usepackage{subfigure}
\usepackage{color}
\usepackage{pifont}
\usepackage{booktabs}
\usepackage{multirow}
\usepackage[utf8]{inputenc}
\usepackage{hyperref}

\newcommand{\ie}{\textit{i.e.}~}

\newcommand{\etal}{\textit{et al.}~}
\newcommand{\Fig}[1]{Fig.~\ref{fig:#1}}
\newcommand{\Sec}[1]{Sec.~\ref{sec:#1}}
\newcommand{\Tab}[1]{Table~\ref{tab:#1}}
\newcommand{\LVendo}{LV\textsubscript{Endo}}
\newcommand{\LVepi}{LV\textsubscript{Epi}}
\newcommand{\LVef}{LV\textsubscript{EF}}
\newcommand{\LVedv}{LV\textsubscript{EDV}}
\newcommand{\LVesv}{LV\textsubscript{ESV}}
\newcommand{\lunet}{\mbox{LU-Net}}
\newcommand{\unet}{\mbox{U-Net}}
\newcommand{\unetone}{\mbox{U-Net1}}
\newcommand{\sota}{\mbox{state-of-the-art}}

\definecolor{orange}{rgb}{1.0, 0.5, 0.0}
\definecolor{mygreen}{rgb}{0.0, 0.5, 0.0}
\newcommand \modified[1]{\textcolor{black}{#1}}

\begin{document}

\title{LU-Net: a multi-task network to improve the robustness of \modified{segmentation of left ventriclular structures by deep learning} in 2D echocardiography}


\author{Sarah~Leclerc, 
		Erik~Smistad,
        Andreas~{\O}stvik,
        Frederic Cervenansky,
        Florian~Espinosa,
        Torvald~Espeland,
        Erik~Andreas~Rye~Berg,
        Thomas~Grenier,
        Carole~Lartizien,
        Pierre-Marc~Jodoin,
        Lasse~Lovstakken, 
        and Olivier~Bernard
\thanks{S.~Leclerc, T.~Grenier, C.~Lartizien, F.~Cervenansky and O.~Bernard are with the Univ Lyon, INSA‐Lyon, Université Claude Bernard Lyon 1, UJM-Saint Etienne, CNRS, Inserm,
CREATIS UMR 5220, U1206, F‐69621, LYON, France. \mbox{E-mail}: olivier.bernard@creatis.insa-lyon.fr.}
\thanks{E.~Smistad, A.~Ostvik and L.~Lovstakken are with the Center of Innovative Ultrasound Solutions, Department of Circulation and Medical Imaging, Norwegian University of Science and Technology, Trondheim, Norway}
\thanks{F.~Espinosa is with the Cardiovascular department Centre Hospitalier de Saint-Etienne
Saint-Etienne, France}
\thanks{T.~Espeland and E.A.~Rye~Berg are with the Center of Innovative Ultrasound Solutions and the Clinic of cardiology, St. Olavs Hospital, Trondheim, Norway}
\thanks{P.-M.~Jodoin is with the Computer Science Department, University of Sherbrooke, Sherbrooke, Canada.}
\vspace{-1cm}}

\markboth{arXiv version}%
{Shell \MakeLowercase{\textit{et al.}}: Bare Demo of IEEEtran.cls for IEEE Journals}

\maketitle

\begin{abstract}
Segmentation of cardiac structures is one of the fundamental steps to estimate volumetric indices of the heart. This step is still performed semi-automatically in clinical routine, and is thus prone to inter- and intra-observer variability. Recent studies have shown that deep learning has the potential to perform fully automatic segmentation. However, the current best solutions still suffer from a lack of robustness. In this work, we introduce an end-to-end multi-task network designed to improve the overall accuracy of cardiac segmentation while enhancing the estimation of clinical indices and reducing the number of outliers. Results obtained on a large open access dataset show that our method outperforms the current best performing deep learning solution and achieved an overall segmentation accuracy lower than the intra-observer variability for the epicardial border (\ie on average a mean absolute error of $1.5$~mm and a Hausdorff distance of $5.1$~mm) with $11$\% of outliers. Moreover, we demonstrate that our method can closely reproduce the expert analysis for the end-diastolic and end-systolic left ventricular volumes, with a mean correlation of $0.96$ and a mean absolute error of $7.6$~ml. Concerning the ejection fraction of the left ventricle, results are more contrasted with a mean correlation coefficient of $0.83$ and an absolute mean error of $5.0$\%, producing scores that are slightly below the intra-observer margin. Based on this observation, areas for improvement are suggested.

\end{abstract}


\IEEEpeerreviewmaketitle

\section{Introduction}
\label{sec:introduction}

Analysis of 2D echocardiographic images based on the measurement of cardiac morphology and function is essential for diagnosis. Low-level image processing such as segmentation and tracking enable to extract and interpret clinical indices, among which the volume of the left ventricle and the corresponding ejection fraction (\LVef) are among the most commonly used. The extraction of such measures requires accurate delineation of the left ventricular endocardium (\LVendo) at both end diastole (ED) and end systole (ES). However, these indices are subject to controversy due to a lack of reproducibility. Indeed, there is a significant variability in the measurement of the values extracted from the ultrasound images from an inter-expert, intra-expert and inter-equipment perspective. The inherent difficulties for segmenting echocardiographic images are well documented: \emph{i)} poor contrast between the myocardium and the blood pool; \emph{ii)} brightness inhomogeneities; \emph{iii)} variation in the speckle pattern along the myocardium, due to the orientation of the cardiac probe with respect to the tissue; \emph{iv)} presence of trabeculae and papillary muscles with intensities similar to the myocardium; \emph{v)} significant tissue echogenicity variability within the population; \emph{vi)} shape, intensity and motion variability across patients and pathologies.

Numerous studies have been conducted for more than 30 years to make automatic measurements of the \LVendo~and \LVef~indices robust and reliable in echocardiographic imaging. In order to achieve an objective evaluation and comparison of \sota~methods, open-access references datasets are essential. In the context of left ventricle analysis, Bernard \etal \cite{Bernard2016} published a dataset composed of 45 sequences of 3D echocardiographic images in conjunction with the Challenge on Endocardial Three-dimensional Ultrasound Segmentation (CETUS), which took place during the MICCAI 2014 conference\footnote{https://www.creatis.insa-lyon.fr/Challenge/CETUS/}. This study, along with additional recent works~\cite{Pedrosa2017,Oktay2018}, revealed that the BEAS approach (deformable contour based on an explicit representation of the evolving surface through a B-spline formalism~\cite{Pedrosa2017}) currently provides the best scores in terms of segmentation of the 3D left endocardium surface and the corresponding \LVendo~and \LVef~estimation but that these results are still higher than the inter-observer variability measured from the same dataset. 

Despite the fact that the training dataset of CETUS consists of only 15 patients, machine learning methods, and especially deep learning methods, produce results that are very close to the best performing ones \cite{Oktay2018}. This led to a recent work whose aim was to study the performance of convolutional neural networks (CNNs) methods for the segmentation of 2D echocardiographic sequences from a larger dataset \cite{Leclerc_tmi_2019}. In particular, the authors set up an open access dataset, named CAMUS, composed of two and four-chamber acquisitions of 2D echocardiographic sequences from 500 patients with reference measurements from one cardiologist on the full dataset and from three cardiologists on a fold of 50 patients. This study revealed that approaches based on encoder-decoder architecture, in particular the well-known U-Net method~\cite{Ronneberger2015}, produce accurate results that are much better than the \sota, on average lower than the inter-observer variability and close but still above the intra-observer variability. Thus, deep learning methods appear to faithfully reproduce the experts' annotations in echocardiographic image segmentation. In this context, the purpose of this paper is to provide answers to the following three questions:
\begin{enumerate}
\item Is it possible to further improve the accuracy of CNNs for the segmentation in echocardiographic imaging?
\item Can the number of outliers be significantly reduced? 
\item Do CNNs allow the achievement of results below intra-observer variability both in terms of segmentation and clinical index estimation?
\end{enumerate}

\section{Previous work}
\label{previous_work}

The study conducted in~\cite{Leclerc_tmi_2019} highlighted two interesting outputs: \emph{i)} the scores produced by \unet~models are not much sensitive to the choice of hyper-parameters, which reinforces the quality of the results obtained by such architecture; \emph{ii)} the use of more sophisticated encoder-decoder architectures (\ie \mbox{U-Net++}~\cite{zhou2018}, stacked hourglasses network~\cite{newell2016} and anatomically constrained neural network~\cite{Oktay2018}) did not produce better results. Therefore, while \unet~ appears as a good choice for the segmentation of echocardiographic images, the improvement of its performance through the extension of its architecture is not straightforward. 

In parallel, there has been an increasing interest in the computer vision community for methods based on attention-learning
to improve classification~\cite{cvpr_wang_2017}, localization~\cite{nips_ren_fast_r_cnn_2015,cvpr_redmon_2016} and segmentation tasks~\cite{iccv_he_2017}. Attention learning correspond to the set of approaches which integrate a contextualization procedure inside their pipeline to improve their overall performance. Contextualization is usually applied either on the image itself or on a derived feature space. One of the best performing approaches is the Mask R-CNN method recently proposed by He \textit{et al.}, which provides the best current results in all three tracks of the COCO suite of challenges \cite{iccv_he_2017}. This network is mainly composed of three stages: \emph{i)} a region proposal network (RPN) which scans boxes distributed over the image area and finds the ones that contain objects; \emph{ii)} a classification network that scans each of the regions of interest (ROIs) proposed by the RPN and assigns them to different classes while refining the location and size of the bounding box to encapsulate the object; \emph{iii)} a convolutional network that takes the regions selected by the ROIs classifier and generates masks for them.

Attention-based approaches have also been successfully applied in medical imaging~\cite{stacom_payer_2017,prl_guan_2018,miccai_wang_2018,oktay_midl_2018,stacom_li_2019,media_vigneault_2018,media_pesce_2019,media_schlemper_2019}. In~\cite{media_vigneault_2018}, the authors proposed a dedicated CNN architecture for simultaneous localization and segmentation in cardiac MR imaging. Their model is built around three stages: \emph{i)} an initial segmentation is performed on the input image; \emph{ii)} the features learned at the bottom layers are then used to predict the parameters of a spatial transformer network that transforms the input image into a canonical orientation~\cite{nips_jaderberg_2015}; \emph{iii)} a final segmentation is performed on the transformed image. In parallel, two attention learning networks were developed in~\cite{media_pesce_2019} for the detection of chest radiographs containing pulmonary lesions. The annotated lesions were used during the training process to deliver visual attention feedback informing the networks about their lesion localisation performance. The first network extracts saliency maps from high level layers and compares the predicted position of a lesion with the true position. The second approach consists of a recurrent attention model which learns to process a short sequence of smaller image portions. Recently, a generic attention model was proposed to automatically learn to focus on target structures in medical image analysis~\cite{media_schlemper_2019}. Based on attention gate modules that can be integrated in any existing CNN architecture~\cite{oktay_midl_2018}, the proposed formalism intrinsically promotes the suppression of irrelevant regions in an input image while highlighting salient features useful for a specific task. This approach has been evaluated for 2D fetal ultrasound image classification and 3D-CT abdominal image segmentation. 

Despite their established interest, to our knowledge, only one study based on an attention model has been conducted so far in echocardiographic image segmentation. In particular, the authors introduced an attention mechanism based on the multiplication of a contextualization map derived from a first network with the input image in order to provide as input a pre-processed image without irrelevant information to a second \unet~network that performs the segmentation of both the left ventricle and the myocardium simultaneously ~\cite{leclerc_ius_2019}. Results on the CAMUS dataset show that this method allows for a reduction of outliers in terms of segmentation results (from 20\% to 17\%) while preserving the same level of accuracy.

\section{Methodology}
\label{sec:methodology}

Since CAMUS is the current largest open access 2D echocardiographic dataset with an active evaluation website, we made the choice to build our study on this dataset. Based on the literature review carried out in the previous section, we decided to investigate the capacity of attention-based networks to improve the current best segmentation scores in 2D echocardiographic imaging.

\subsection{Motivations}
\label{sec:motivations}

The work carried out in this study was motivated by an experiment we conducted on the CAMUS dataset, whose details are described below. In particular, we manually selected regions of interest (ROIs) around the reference segmentation masks. Each ROI corresponds to the ideal bounding box (BB) surrounding the corresponding mask with an additional margin $m$ of $5$, $15$ and $30$\% along the axes. From these ROIs, the corresponding images were cropped to create new datasets that were processed with the baseline \mbox{\unet1} architecture described in~\cite{Leclerc_tmi_2019}. The corresponding scores are reported in \Tab{segmentation_good_medium} and referred to as \mbox{BB-m5}, \mbox{BB-m15} and \mbox{BB-m30}, respectively. From this table, it is worth noting the contribution of the cropping stage, leading to a significant improvement of the baseline \unetone~results, with average scores all below the ones of the intra-observer (except for \mbox{BB-m30} with the Hausdorff distance metric) and a number of outliers lower than $8$\%. This experiment thus reveals that the effective insertion of a localization step during the segmentation process with the \unet~architecture would yield remarkable results in echocardiographic image segmentation.

\subsection{Overall strategy}
\label{sec:overall_strategy}

\begin{figure*}[!thp]
\centerline{\includegraphics[height= 8cm]{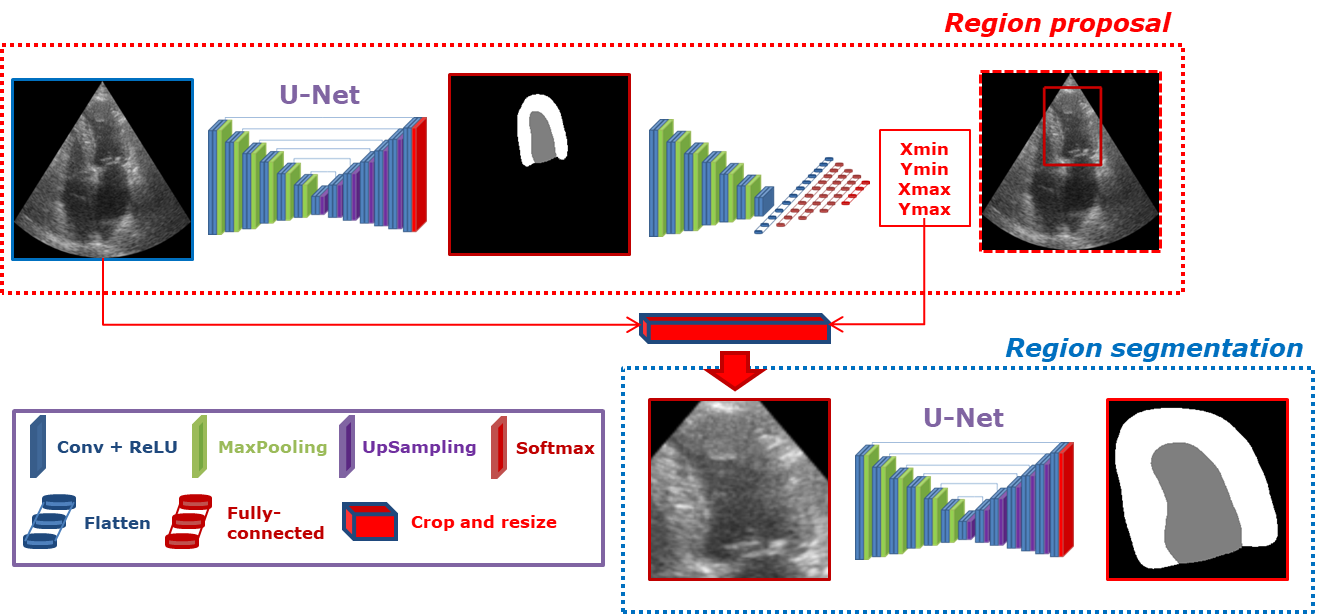}}
\caption{Illustration of the \lunet~pipeline with the U-L2-mu localization network introduced in \Sec{localization_part}. The two U-Nets are independent.}
\label{fig:lunet}
\end{figure*}

Based on the motivations and the literature review on attention learning presented in the previous sections, we developed a multi-task network to improve the robustness of segmentation in 2D echocardiography. Since the \unet~model already produces high-performance segmentation results in echocardiography~\cite{Leclerc_tmi_2019}, we decided to use this architecture as back-bone for our multi-task network, referred to as Localization \unet~(\lunet) in the sequel. \lunet~aims at locating and segmenting the endocardial and the epicardial borders of the left ventricle through an end-to-end learning procedure. The underlying assumption of this strategy is that the joint optimization of these two tasks should lead to better segmentation results. An illustration of the \lunet's overall architecture is provided in \Fig{lunet}. In particular, \lunet~is composed of two networks: one region proposal network for localization and one \unet~for segmentation. 

\vspace*{0.3cm}
\subsubsection{Localization part}
\label{sec:localization_part}

The region proposal network performs a mapping between the input ultrasound image and four coordinates to define a bounding box (BB) around the structure of interest, namely the union of the left ventricle and myocardium. The reference BB is defined as the minimal bounding box in contact with the epicardium border. The target coordinates are computed with an additional margin $m$ as:
\begin{align*}
    x^m_{min} &= x_{min} - m * h  \quad \text{and} \quad x^m_{max} = x_{max} + m * h, \notag \\
    y^m_{min} &= y_{min} - m * w  \quad \text{and} \quad y^m_{max} = y_{max} + m * w.
\end{align*}
where $\left(x_{min},x_{max},y_{min},y_{max}\right)$ are the coordinates of the reference BB and $(w,h)$ its width and height. The motivation for adding a margin was to provide some context around the targeted structures for the segmentation task.

\vspace*{0.2cm}
\subsubsection{Segmentation part}
\label{sec:segmentation_part}

The output of the region proposal network is used as an attention mechanism to crop and resize the input ultrasound image. The resulting image is fed to a segmentation network which corresponds to a \unet (the \unetone~model described in~\cite{Leclerc_tmi_2019}), currently the most efficient model evaluated on the CAMUS dataset considering a trade-off between accuracy, speed and size. 

\vspace*{0.2cm}
\subsubsection{End-to-end approach}
\label{sec:end_to_end_part}

In order to make the full network trainable end-to-end, the crop and resize step was implemented using a bilinear differentiable sampling. In addition, the segmentation loss involved in the second \unet~was modified to evolve dynamically over the training phase with respect to the varying ROI. The two U-Nets are independent networks with distinct parameters. At inference time, based on the localization outputs, the final segmentation result is then returned to the original coordinate system of the input image.

\section{Experiments}
\label{sec:experiments}

\subsection{Dataset}
\label{sec:dataset}

The CAMUS dataset contains two and four-chamber acquisitions from 500 patients~\cite{Leclerc_tmi_2019}. The full dataset was divided into 10 folds equally distributed in terms of image quality (good, medium, poor) and ejection fraction category ($\leq$ 45\%, $\geq$ 55\% or in between). This allows the analysis of the full dataset by means of a classical cross-validation strategy. One cardiologist \modified{(\emph{O\textsubscript{1}})} manually annotated the endocardium and epicardium (\LVepi) borders of the left ventricle on the full dataset at end diastole (ED) and end systole (ES) and two other cardiologists \modified{(\emph{O\textsubscript{2}} and \emph{O\textsubscript{3}})} on a fold of 50 patients. \modified{This fold was also annotated twice by \emph{O\textsubscript{1}} seven months apart. This procedure} allows comparison of the results provided by the algorithms with the inter- and intra-observer variability.

\subsection{Evaluation metrics}
\label{sec:evalaution_metrics}

\subsubsection{Localization metrics}
\label{sec:localization_metrics}

We assessed the performance of the localization networks through the Intersection Over Union (IOU) metric and the euclidean distance errors between the predicted and the reference BB coordinates (\ie its central position $\left(x_c,y_c\right)$, its height $h$ and width $w$). The IOU is a classical localization metric which measures the overlap between the predicted BB and the reference one. It gives a value between 0 (no overlap) and 1 (full overlap). In addition, we provided the "BB out" metric which corresponds to the number of cases where the predicted BB does not completely encompass the reference mask.

\vspace*{0.2cm}
\subsubsection{Segmentation metrics}
\label{sec:segmentation_metrics}

To measure the accuracy of the segmentation output (\LVendo~and \LVepi) of a given method, the Dice metric (closely related to the IOU and classically used in segmentation), the mean absolute distance ($d_{m}$) and the 2D Hausdorff distance ($d_H$) were used. The Dice similarity index is a measure of overlap between the segmented surface $S_{user}$ extracted from a method and the corresponding reference surface $S_{ref}$. It gives a value between 0 (no overlap) and 1 (full overlap). $d_{m}$ corresponds to the average distance between $S_{user}$ and $S_{ref}$ while $d_H$ measures the maximum local distance between the two surfaces. In addition, we assessed the quality of segmentation with regard to cardiologists' annotations through the notion of outliers defined bellow.
\begin{itemize}
    \item geometric outlier: the set of segmentation attached to a patient is seen as a geometric outlier if at least one of its eight corresponding distance scores (\ie $d_{m}$ and $d_{H}$ values at ED and ES for both apical two and four-chamber views) is out of the corresponding bounds defined from the inter-observer variability~\cite{Leclerc_tmi_2019}; 
    \item anatomical outlier: the set of segmentation attached to a patient is seen as an anatomical outlier if the simplicity and convexity ~\cite{leclerc_ius_2019} of the corresponding segmented contours are lower than the lowest values computed from expert annotations on $50$ patients. \modified{These two metrics hold values between 0 and 1, and are maximized for a circle. They also give discriminating values for any convex shapes, such as oval shapes like heart cavities, and bridge-like shapes like the myocardium. They can therefore be used as simple tools to detect anatomical outliers in the case of left ventricular structures.}
\end{itemize}

\vspace*{0.2cm}
\subsubsection{Clinical metrics}
\label{sec:geometric_metrics}

We evaluated the performance of the methods with $3$ clinical indices: \emph{i)} the ED volume (\LVedv~in $ml$); \emph{ii)} the ES volume (\LVesv~in $ml$); \emph{iii)} the ejection fraction (\LVef~as a percentage), for which we computed two metrics: the correlation ($corr$) and the limit of agreement ($loa$) (computed from conventional definitions). \modified{Please note that all volumes of the left ventricle were computed using Simpson’s biplane rule \cite{Folland1979} involving the segmentation results on the two- and four-chamber apical views}

\subsection{Localization methods}
\label{sec:localization_methods}

We implemented and assessed the performance of four different convolutional networks dedicated to the prediction of bounding boxes, 
\ie predicting $\left(x^m_{min},x^m_{max},y^m_{min},y^m_{max}\right)$. 


\vspace*{0.2cm}
\begin{enumerate}
    \item an AlexNet-like network \cite{Krizhevsky_nips_2012} composed of a succession of convolutional layers of varying filter size and max pooling. Our version ends with three fully-connected layers of size 4096, 4096 and 4. Except for the additional last layer, this architecture is therefore the same as the original, without any dropout and data augmentation strategy, and includes 71M parameters;
    \item a VGG19-like network \cite{Simonyan2014}, composed of 20 layers that alternate between convolutions and max pooling. The last fully-connected layers are made respectively of 4096, 4096 and 4 units, for a total of 70M parameters;
    \item U-L1 based on a \unet~model performing the segmentation of the left ventricle and the myocardium. The bottom layer of this \unet~was derived in order to carry out the localization procedure using four fully connected layers of 1024, 256, 32, and 4 units. This model was inspired by the work of Vigneault~\etal \cite{media_vigneault_2018}. The network includes 9M parameters;
    \item U-L2 also based on a \unet~model performing the segmentation of the left ventricle and the myocardium. The output of this \unet~was then connected to a downsampling branch ending with four fully-connected layers of 1024, 256, 32 and 4 units. This architecture is novel and corresponds to one of the innovations proposed in this study. We evaluated two versions of this network, one optimizing only the localization loss (referred to as \mbox{U-L2-mo}) and one optimizing both the localization and the segmentation losses (referred to as \mbox{U-L2-mu}). The network includes 11M parameters.
\end{enumerate}

\subsection{Segmentation methods}
\label{sec:segmentation_methods}

The performance of the joint segmentation of the endocardial and the epicardial borders was assessed through the following four networks:

\vspace*{0.2cm}
\begin{enumerate}
    \item \unetone, corresponding to the current best performing network on the CAMUS dataset~\cite{Leclerc_tmi_2019}. This network includes 2M parameters;
    \item RU-Net, recently introduced in~\cite{leclerc_ius_2019} and built from two cascaded \unetone. The epicardial mask predicted by the first network is dilated and multiplied with the input image to provide a contextualized image as input to the second network, with a total number of 4M parameters;
    \item Attention-gated U-Net (\mbox{AG-U-Net}), recently proposed in~\cite{oktay_midl_2018}, in which attention layers are used at each skip connection to locally weigh the concatenated features with coefficients derived from the previous layer. It includes batch normalization before each activation, and deep supervision by aggregating the feature maps produced after each attention layer at the last level of \unetone (\ie before the last convolution and the softmax). This network has a total number of 2M parameters;
    \item LU-Net, as introduced in this paper, built using \mbox{U-L2-mu} as the region proposal network and \unetone~as the segmentation network for a total of 13M parameters. Two margins of $m=5$\% and $m=15$\% were evaluated.
\end{enumerate}

\subsection{Learning strategy}
\label{sec:learning}

\subsubsection{Optimizer}

All the methods involved in this study were optimized using Adam's optimizer associated to a learning rate (either equal to $1e^{-3}$ or $1e^{-4}$) and a number of epochs (controlled using early stopping with a patience of 20) that experimentally allowed to observe a smooth convergence of the training and validation losses. The best model on the validation loss was selected after each training phase.  

\vspace*{0.2cm}
\subsubsection{Loss}

Localization networks were optimized using a L1 loss clipped at $0.99$ summing the errors on the four BB values (\ie $\left(x^m_{min},x^m_{max},y^m_{min},y^m_{max}\right)$). Segmentation networks were optimized using a multi-class Dice loss taking into account the LV and myocardium predictions. For multi-task prediction, a weighting of 10 was applied to the localization term as to balance the localization and the segmentation objectives.

\section{Results}
\label{sec:results}

In order to easily compare our results with those of the \sota~on the CAMUS dataset, we followed the strategy developed in \cite{Leclerc_tmi_2019} by training for each deep learning method a single model on the annotated images of both apical two and four-chamber views, regardless of the time instant. 

\subsection{Localization results}
\label{sec:localization_results}

\Tab{localization_results} shows the localization accuracy computed on the full dataset ($500$ patients) for the algorithms described in section \Sec{experiments}. Mean and standard deviation values for each metric were obtained from cross-validation on the 10 folds of the dataset (see~\cite{Leclerc_tmi_2019} for more details). For each row of this table, the \emph{m} information indicated after the name of the method indicates the margin value used to define the reference BB. The values in bold correspond to the best scores for each metric. 

Based on a comparison of the methods using a margin of 5\%, the proposed \mbox{U-L2-mu} gets the overall best localization scores on all metrics, except for the error on $y_c$ with a difference of 0.2 mm with the best method. These results validate the use of the \unet~architecture, which has already proven its effectiveness in terms of segmentation, to perform localization tasks in ultrasound imaging compared to well-established computer vision architectures (\ie AlexNet and VGG). In addition, the scores highlight the interest of using both segmentation and localization losses to improve the performance of the \mbox{U-L2} method, with an average gain of $2.5$~mm over the BB centre estimate and $3.6$~mm over the BB dimension estimate. This significant improvement demonstrates that forcing segmentation as an intermediate step to localization is beneficial.

We also investigated the influence of the choice of the margin value \emph{m} on the accuracy of the localization results produced by the \mbox{U-L2} method. The obtained results are contrasted. Indeed, while the use of a lower margin (\ie 5\%) produces slightly better results with regard to the estimation of the BB position, the use of a higher margin (\ie 15\%) considerably reduces the number of cases where the BB does not encompass the reference mask (from 36\% to 2\%).

Based on this experiment, it is clear that the \mbox{U-L2-mu} model produced the best localization results. We therefore decided to use this network as the region proposal part of the \lunet~architecture, as illustrated in \Fig{lunet}.

\begin{table}[t!]
     \caption{Localization accuracy on 4 evaluated methods on the full dataset (500 patients). The $m$ information contained in each method name indicates the margin value defined in \Sec{localization_part}}
 \begin{center}
\begin{tabular}{c c *{5}{c}}
 \toprule 
 \multicolumn{1}{l}{\multirow{2}{*}{\bf \small Model}} &
\multicolumn{1}{c}{ \multirow{2}{*}{\bf \small IOU}} & \multicolumn{4}{c}{\bf{Error (mm)}} & \bf{BB out} \\
\cmidrule(r){3-6}
 & & $x_c$ & $y_c$ & h & w &  \\
\midrule
\multicolumn{1}{l}{\multirow{2}{*}{AlexNet-m5}} & 0.880 & 2.2 &  1.9 & 4.2 & 4.1 & 866\\
& \scriptsize{$\pm{0.062}$} & \scriptsize{$\pm{2.4}$} & \scriptsize{$\pm{1.8}$} & \scriptsize{$\pm${4.1}} & \scriptsize{$\pm${4.1}} & \scriptsize{43\%}\\
\multicolumn{1}{l}{\multirow{2}{*}{VGG-m5}} & 0.888 & 1.9 & \bf 1.7 & 4.0 & 4.0 & 903\\
& \scriptsize{$\pm{0.060}$} & \scriptsize{$\pm{2.4}$} & \scriptsize{$\pm{1.7}$} & \scriptsize{$\pm${3.9}} & \scriptsize{$\pm${3.7}} & \scriptsize{45\%}\\
\multicolumn{1}{l}{\multirow{2}{*}{U-L1-m5}} & 0.849 & 3.1 & 2.7 & 5.3 & 4.9 & 1094\\
& \scriptsize{$\pm{0.072}$} & \scriptsize{$\pm{2.9}$} & \scriptsize{$\pm{2.4}$} & \scriptsize{$\pm${4.5}} & \scriptsize{$\pm${4.3}} & \scriptsize{55\%}\\
\multicolumn{1}{l}{\multirow{2}{*}{U-L2-mo-m5}} & 0.791 & 4.2 &  4.4 & 7.1 & 6.9 & 1393\\
& \scriptsize{$\pm{0.138}$} & \scriptsize{$\pm{4.7}$} & \scriptsize{$\pm{6.0}$} & \scriptsize{$\pm${6.4}} & \scriptsize{$\pm${6.7}} & \scriptsize{70\%}\\
\multicolumn{1}{l}{\multirow{2}{*}{U-L2-mu-m5}} & 0.898 & \bf 1.6 & 1.9 & \bf 3.2 & \bf 3.6 & 712\\
& \scriptsize{$\pm{0.053}$} & \scriptsize{$\pm{1.8}$} & \scriptsize{$\pm{1.9}$} & \scriptsize{$\pm${3.1}} & \scriptsize{$\pm${3.2}} & \scriptsize{36\%}\\
\midrule
\multicolumn{1}{l}{\multirow{2}{*}{U-L2-mu-m15}} & \bf 0.907 & \bf 1.6 & \bf 1.7 & 3.7 & 4.3 & \bf 31\\
& \scriptsize{$\pm{0.054}$} & \scriptsize{$\pm{2.0}$} & \scriptsize{$\pm{1.7}$} & \scriptsize{$\pm${4.0}} & \scriptsize{$\pm${4.3}} & \scriptsize{2\%}\\
\bottomrule
   \label{tab:localization_results}
  \end{tabular}
  \end{center}
\end{table}

\subsection{Segmentation results}
\label{sec:segmentation_results}

\begin{figure}[tp]
\centering
\subfigure[]{\includegraphics[width=4cm]{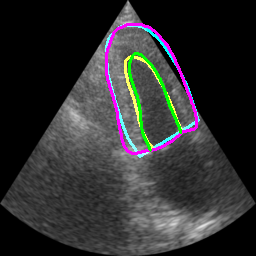} \hspace{2mm} \includegraphics[width=4cm]{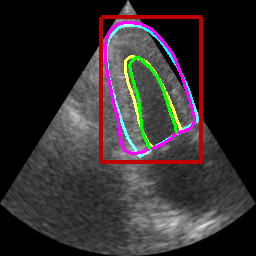}}%
\vfil
\subfigure[]{\includegraphics[width=4cm]{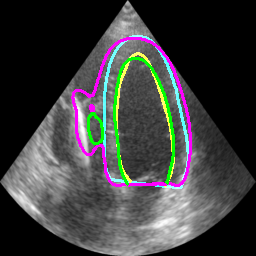} \hspace{2mm} \includegraphics[width=4cm]{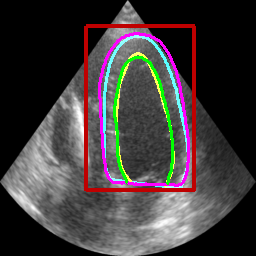}}%
\vfil
\subfigure[]{\includegraphics[width=4cm]{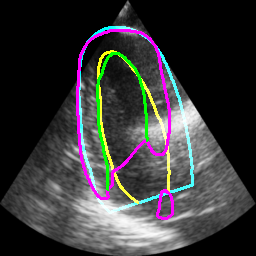} \hspace{2mm} \includegraphics[width=4cm]{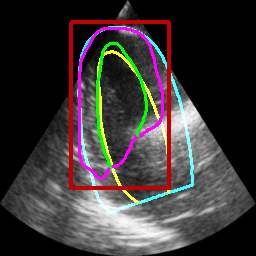}}%
\caption{Comparison of the segmentation performance of the baseline \unetone~(left column) and the proposed \lunet~architecture (right column) on cases (a) with similar results; (b) where the intermediate localization of the \lunet~helps; (c) where the artifact present in the image is too strong for any improvement. In each image, the prediction is in green and purple while the ground-truth is in yellow and cyan. The BB estimated is displayed in red.}
\label{fig:visuals}
\end{figure}

\Tab{segmentation_good_medium} displays the segmentation accuracy computed on the full dataset from patients having good and medium image quality ($406$ patients) for the 4 algorithms described in section \Sec{segmentation_methods}. Mean and standard deviation values for each metric were obtained from cross-validation on the 10 folds of the dataset. The values in bold correspond to the best scores for each metric. From these results, one can see that all the attention-based networks produced either the same, or better results than the baseline \unetone, with \mbox{AG-U-Net} and \lunet~being the best performing models.
Indeed, \mbox{AG-U-Net} obtained the overall best results for the segmentation of the \LVendo~border ($d_m$ value of $1.5$~mm and $d_H$ value of $5.3$~mm), leading to segmentation scores close but still higher than the intra-observer variability for this structure. The \mbox{LU-Net-m5} approach obtained the best results for the segmentation of the \LVepi~border ($d_m$ value of $1.5$~mm and $d_H$ value of $5.1$~mm) and the lowest number of geometric outliers ($11$\%). Interestingly, these scores are either equivalent or lower than the intra-observer variability for this structure. It is also worth noting the robustness of the \lunet~model with respect to the choice of margin parameter, as margins of $m=5$\% and $m=15\%$ produce almost the same segmentation scores for all metrics. An illustration of the segmentation performance of the \mbox{LU-Net-m5} network compared to the baseline \unetone~model on three different cases is provided in \Fig{visuals}.


\setlength{\tabcolsep}{1.0em}
\begin{table*}[t!]
     \caption{Segmentation accuracy on the 4 evaluated methods described in \Sec{segmentation_methods} and restricted to patients having good and medium image quality (406 in total). The $m$ information contained in each methods name indicates the margin value defined in \Sec{localization_part}}
 \begin{center}
\begin{tabular}{c c*{7}{c}}
 \toprule 
 & \multicolumn{1}{l}{\multirow{4}{*}{\bf \small Model}} &
 \multicolumn{3}{c}{\bf{\LVendo}} & \multicolumn{3}{c}{\bf{\LVepi}} & \bf{outliers} \\
 \cmidrule(r){3-5} \cmidrule(r){6-8}
 & & \multicolumn{1}{c}{\bf \emph{D}} & \multicolumn{1}{c}{\bf \emph{d\textsubscript{m}}} & \bf \emph{d\textsubscript{H}} & \bf \emph{D} & \bf \emph{d\textsubscript{m}} & \bf \emph{d\textsubscript{H}} & \bf geo. \\
  \cmidrule(r){3-3} \cmidrule(r){4-4} \cmidrule(r){5-5} \cmidrule(r){6-6} \cmidrule(r){7-7} \cmidrule(r){8-8} 
 & & val. & mm & mm & val. & mm & mm &  \# \%\\
\midrule

& \multicolumn{1}{l}{\multirow{2}{*}{intra-observer}} &  0.937 & 1.4 & 4.5  & 0.954 & 1.7 & 5.0 & 21\\
& & \scriptsize{$\pm${0.027}} & \scriptsize{$\pm${0.5}} & \scriptsize{$\pm${1.8}} & \scriptsize{$\pm${0.020}} & \scriptsize{$\pm${0.8}} & \scriptsize{$\pm${2.2}} & \scriptsize{13}\\
 \midrule
 \multirow{6}{*}{\rotatebox{90}{\parbox{1.3cm}{Motivation \\ study (\Sec{motivations})}}} & \multicolumn{1}{l}{\multirow{2}{*}{BB-m5}} & 0.941 & 1.3 & 4.3  & 0.971 & 1.0 & 4.1 & 89\\
 & & \scriptsize{$\pm${0.034}} & \scriptsize{$\pm${0.6}} & \scriptsize{$\pm${1.9}} & \scriptsize{$\pm${0.011}} & \scriptsize{$\pm${0.4}} & \scriptsize{$\pm${1.8}} & \scriptsize{5.5}\\
 
 & \multicolumn{1}{l}{\multirow{2}{*}{BB-m15}} & 0.940 & 1.3 & 4.4  & 0.969 & 1.1 & 4.3 & 106\\
 & & \scriptsize{$\pm${0.034}} & \scriptsize{$\pm${0.6}} & \scriptsize{$\pm${1.9}} & \scriptsize{$\pm${0.011}} & \scriptsize{$\pm${0.4}} & \scriptsize{$\pm${2.0}} & \scriptsize{6.5}\\
 
 & \multicolumn{1}{l}{\multirow{2}{*}{BB-m30}} & 0.937 & 1.4 & 4.7  & 0.966 & 1.2 & 4.6 & 124\\
 & & \scriptsize{$\pm${0.035}} & \scriptsize{$\pm${0.6}} & \scriptsize{$\pm${2.1}} & \scriptsize{$\pm${0.013}} & \scriptsize{$\pm${0.5}} & \scriptsize{$\pm${2.2}} & \scriptsize{7.6}\\
 \midrule

\multirow{10}{*}{\rotatebox{90}{\parbox{1.5cm}{Experimental \\ study (\Sec{segmentation_results})}}} & \multicolumn{1}{l}{\multirow{2}{*}{\unetone}} & 0.920 & 1.7 & 5.6  & 0.947 & 1.9 & 6.2 & 282 \\
& & \scriptsize{$\pm{0.056}$} & \scriptsize{$\pm{1.2}$} & \scriptsize{$\pm{3.3}$} & \scriptsize{$\pm${0.030}} & \scriptsize{$\pm${1.1}} & \scriptsize{$\pm{3.7}$} & \scriptsize{17\%}\\ 

& \multicolumn{1}{l}{\multirow{2}{*}{RU-Net~\cite{leclerc_ius_2019}}} & 0.925 & 1.7 & 5.4  & \bf 0.950 & 1.8 & 5.8 & 240\\
& & \scriptsize{$\pm${0.049}} & \scriptsize{$\pm${1.0}} & \scriptsize{$\pm${3.3}} & \scriptsize{$\pm${0.030}} & \scriptsize{$\pm${1.1}} & \scriptsize{$\pm${3.9}} & \scriptsize{15\%}\\
 
& \multicolumn{1}{l}{\multirow{2}{*}{AG-U-Net~\cite{oktay_midl_2018}}} & 0.930 & \bf 1.5 & \bf 5.3  & \bf 0.950 & 1.8 & 5.9 & 270\\
& & \scriptsize{$\pm${0.049}} & \scriptsize{$\pm${1.3}} & \scriptsize{$\pm${3.4}} & \scriptsize{$\pm${0.026}} & \scriptsize{$\pm${1.0}} & \scriptsize{$\pm${3.7}} & \scriptsize{17\%}\\
 
& \multicolumn{1}{l}{\multirow{2}{*}{LU-Net-m5}} & \bf{0.953} & 1.7 & 5.5  & 0.932 & \bf 1.5 & \bf 5.1 & \bf 186\\
& & \scriptsize{$\pm${0.026}} & \scriptsize{$\pm${0.9}} & \scriptsize{$\pm${3.6}} & \scriptsize{$\pm${0.043}} & \scriptsize{$\pm${0.8}} & \scriptsize{$\pm${3.3}} & \scriptsize{11\%}\\
& \multicolumn{1}{l}{\multirow{2}{*}{LU-Net-m15}} & 0.952 & 1.7 & 5.6 & 0.931 & \bf 1.5 & 5.3 & 203\\
& & \scriptsize{$\pm${0.029}} & \scriptsize{$\pm${1.1}} & \scriptsize{$\pm${4.0}} & \scriptsize{$\pm${0.049}} & \scriptsize{$\pm${1.1}} & \scriptsize{$\pm${3.6}} & \scriptsize{12\%}\\
\midrule
 
\multicolumn{9}{l}{} \\
\multicolumn{9}{l}{*~\emph{\LVendo: Endocardial contour of the left ventricle; \LVepi: Epicardial contour of the left ventricle}} \\
\multicolumn{9}{l}{~~\emph{D: Dice index; d\textsubscript{m}: mean absolute distance; d\textsubscript{H}: Hausdorff distance}} \\
\multicolumn{9}{l}{~~\emph{The values in bold refer to the best performance for each measure.}}
 
   \label{tab:segmentation_good_medium}
  \end{tabular}
  \end{center}
\end{table*}

\subsection{Clinical scores}

\Tab{clinical} contains the clinical metrics computed on the full dataset from patients having good and medium image quality (406 patients) for the 4 methods described in \Sec{segmentation_methods}. Those indices were computed with the Simpson's \modified {biplane} rule \cite{Folland1979} from the segmentation results \modified {on the two- and four-chamber apical views} of each algorithm. The values in bold represent the best scores for the corresponding index. As for segmentation, the \mbox{AG-U-Net} and \lunet-m5 models obtained the best clinical scores on all the tested metrics (bias was not taken into account since the lowest bias value in itself does not necessarily mean the best performing method). Regarding the estimation of the \LVedv, the two methods produced high correlation scores ($0.956$), small biases ($\pm 1.4$~ml) and reasonable limit of agreements (around $22$~ml) and mean absolute errors (around $8.3$~ml). The \mbox{AG-U-Net} produced the best \LVesv~results with a correlation of $0.962$, while the \lunet-m5 model produced the best \LVef~scores with a correlation of $0.829$. However, even if the scores of \lunet-m5~and \mbox{AG-U-Net} are slightly better than the baseline \unetone~ones, they are still higher that the intra-observer results. This reveals that there is still room for improvement as discussed in~\Sec{discussion}.



\setlength{\tabcolsep}{1.0em}
\begin{table*}[tbp]
  \caption{Clinical metrics of the 4 evaluated methods described in \Sec{segmentation_methods} and restricted to patients having good and medium image quality (406 in total)}
  \centering
  \begin{tabular}{c*{3}{c}|*{3}{c}|*{3}{c}}

 \toprule 
 \multicolumn{1}{l}{\multirow{5}{*}{\small \bf Model}} &  \multicolumn{3}{c}{\bf \small \emph{LV\textsubscript{EDV}}} & \multicolumn{3}{c}{\bf \small \emph{LV\textsubscript{ESV}}} & \multicolumn{3}{c}{\bf \small \emph{LV\textsubscript{EF}}}\\
 \cmidrule(r){2-4} \cmidrule(r){5-7} \cmidrule(r){8-10}
 & \multicolumn{1}{c}{\bf \emph{corr}} & \multicolumn{1}{c}{\bf \emph{loa}} & \multicolumn{1}{c}{\bf \emph{mae}} & \multicolumn{1}{c}{\bf \emph{corr}} & \multicolumn{1}{c}{\bf \emph{loa}} & \multicolumn{1}{c}{\bf \emph{mae}} & \multicolumn{1}{c}{\bf \emph{corr}} & \multicolumn{1}{c}{\bf \emph{loa}} & \multicolumn{1}{c}{\bf \emph{mae}} \\
\cmidrule(r){2-2} \cmidrule(r){3-3} \cmidrule(r){4-4} \cmidrule(r){5-5} \cmidrule(r){6-6} \cmidrule(r){7-7} \cmidrule(r){8-8} \cmidrule(r){9-9} \cmidrule(r){10-10}
 & val. & ml & ml & val.& ml & ml & val. & \% & \% \\
\midrule
\multicolumn{1}{l}{intra-observer} & 0.978 & -2.8$\pm$14.3& 6.2 & 0.981 & -0.1$\pm$11.4 & 4.5 & 0.896 & -2.3$\pm$11.2 & 4.5\\
\midrule
\multicolumn{1}{l}{\unetone} & 0.947 & -8.3$\pm$24.7 & 10.9 & 0.955 & -4.9$\pm$19.4 & 8.2 & 0.791 & -0.5$\pm$15.1 & 5.6 \\
\multicolumn{1}{l}{RU-Net~\cite{leclerc_ius_2019}} & 0.946 & -1.2$\pm$23.9 & 8.9 & 0.949 & 0.3$\pm$19.6 & 7.3 & 0.704 & -2.1$\pm$14.3 & 6.0 \\
\multicolumn{1}{l}{AG-U-Net~\cite{oktay_midl_2018}} & \bf 0.956 & -1.4$\pm$21.9 & \bf 8.1 & \bf 0.962 & 0.6$\pm$17.0 & \bf 6.2 & 0.798 & -2.2$\pm$15.1 & 5.5 \\
\multicolumn{1}{l}{LU-Net-m5} & \bf 0.956 & 1.4 $\pm$21.8 & 8.3 & 0.956 & 1.6$\pm$ 18.0 & 7.0  & \bf 0.829 & -1.5$\pm$13.5 & \bf 5.0 \\
\multicolumn{1}{l}{LU-Net-m15} & 0.952 & 2.4 $\pm$22.9 & \bf 8.1 & \bf 0.962 & 1.8$\pm$16.7 & 6.5  & 0.821 & -1.2$\pm$13.7 & \bf 5.0 \\
\bottomrule
 \multicolumn{10}{l}{} \\

 \multicolumn{10}{l}{*~\emph{corr: Pearson correlation coefficient; loa: limit of agreement; mae: mean absolute error.}} \\
 \multicolumn{10}{l}{~~\emph{The values in bold refer to the best performance for each measure.}}

\end{tabular}
\label{tab:clinical}
\end{table*}


    
\subsection{\lunet~behavior}

From the results given in \Tab{segmentation_good_medium} and \Tab{clinical}, it appears that the \lunet~method outperforms the baseline \unetone~model both in terms of segmentation and clinical indice estimation. Furthermore, it is one of the most effective model, even compared to other attention-based networks. In order to complete the analysis of \lunet, we applied this network to the full dataset (including poor image quality) and studied the generated outliers. The corresponding results obtained with a margin of $m=5\%$ are provided in \Tab{outliers}. The results of the model named \mbox{LU-Net-m5-o1} corresponds to the scores derived from the output of the first \unet~involved in the region proposal network, while the scores of the model named \mbox{LU-Net-m5-o2} corresponds to the scores derived from the final output of the network (\ie the one provided by the second \unet). From this table, one can see that \lunet~outperforms the \unetone~architecture for all the metrics for both \LVendo~and \LVepi~borders when considering all quality of images. Also, the segmentation results produced by \lunet~appear to be remarkably stable when integrating poor image quality images, with a mean difference of $0.1$~mm for $d_m$, $0.2$~mm for $d_H$ and 1\% for the geometric outliers.

Concerning the localization scores, the \mbox{LU-Net-m5} model obtained consistent results with respect to the \mbox{U-L2-mu} best performing method reported in \Tab{localization_results} with an IOU of $0.906$ and $\left(x_c,y_c,h,w\right)$ BB errors of $\left(1.5,1.5,3.3,3.5\right)$~mm, respectively. Coupling this result with the last two lines of \Tab{outliers} which show that the first segmentation is less accurate than a single \unetone, it appears that the segmentation result produced in the region proposal part of the \lunet~is degraded by optimizing the localization procedure, 
which in turn allows for a significant improvement of the final segmentation results compared to the baseline \unetone~model.

Concerning the segmentation scores, \mbox{LU-Net-m5} produced 12\% of geometric outliers, 2\% of anatomical outliers and 1\% of both, showing that 
half of the anatomical outliers are also geometric. Moreover, the geometric outlier rate is lower than the intra-observer variability one computed from a subset of 40 patients with good and medium image quality, which further highlights the quality of the results achieved by \lunet.

\setlength{\tabcolsep}{0.3em}
\begin{table}[t!]
     \caption{Segmentation accuracy and outliers on the full dataset (500 patients) including those with poor image quality}
 \begin{center}
\begin{tabular}{c*{7}{c}}
 \toprule 
 \multicolumn{1}{l}{\multirow{4}{*}{\bf \small Model}}&
 \multicolumn{2}{c}{\bf \LVendo} & \multicolumn{2}{c}{\bf \LVepi} & \multicolumn{3}{c}{\bf outliers} \\
 \cmidrule(r){2-3} \cmidrule(r){4-5} \cmidrule(r){6-8}
 & \multicolumn{1}{c}{\bf \emph{d\textsubscript{m}}} & \multicolumn{1}{c}{\bf \emph{d\textsubscript{H}}} & \multicolumn{1}{c}{\bf \emph{d\textsubscript{m}}} & \multicolumn{1}{c}{\bf \emph{d\textsubscript{H}}} & \bf geo. & \bf ana. & \bf{both} \\
  \cmidrule(r){2-2} \cmidrule(r){3-3} \cmidrule(r){4-4} \cmidrule(r){5-5} \cmidrule(r){6-8}
 & mm & mm & mm & mm &  \multicolumn{3}{c}{\# \%} \\ 
\midrule

 \multicolumn{1}{l}{\multirow{2}{*}{\unetone}} & 2.0 & 6.1 & 2.0  & 6.5 & 423 & 95 & 71\\
& \scriptsize{$\pm{1.2}$} & \scriptsize{$\pm{3.9}$} & \scriptsize{$\pm${1.1}} & \scriptsize{$\pm${4.5}} & \scriptsize{21\%} & \scriptsize{5\%} & \scriptsize{4\%}\\
\midrule



 \multicolumn{1}{l}{\multirow{2}{*}{LU-Net-m5-o1}} & 2.1 & 7.0 & 1.9 & 6.2 & 483 & 201 & 138\\
& \scriptsize{$\pm{1.1}$} & \scriptsize{$\pm{4.7}$} & \scriptsize{$\pm${1.0}} & \scriptsize{$\pm${3.4}} & \scriptsize{24\%} & \scriptsize{10\%} & \scriptsize{7\%}\\

 \multicolumn{1}{l}{\multirow{2}{*}{LU-Net-m5-o2}} & \bf 1.8 & \bf 5.7 & \bf 1.6  & \bf 5.3 & \bf 240 & \bf 31 & \bf 20\\
& \scriptsize{$\pm{1.0}$} & \scriptsize{$\pm{3.6}$} & \scriptsize{$\pm${0.9}} & \scriptsize{$\pm${3.3}} & \scriptsize{12\%} & \scriptsize{2\%} & \scriptsize{1\%}\\



 \bottomrule
  \end{tabular}
    \label{tab:outliers}
  \end{center}
\end{table}

\section{Discussion}
\label{sec:discussion}

\subsection{Attention-based networks}
\label{sec:attention_based_networks}

\Tab{segmentation_good_medium} and \Tab{clinical} underline the ability of attention-based networks to improve the segmentation and the estimation of clinical indexes in 2D echocardiography. These results are even more interesting given that the authors of the original study ~\cite{Leclerc_tmi_2019} had not succeeded in improving the scores of the baseline \unetone~model through more sophisticated architectures. Although \mbox{AG-U-Net} produced the best scores on the \LVendo~and the estimation of the \LVesv~, \lunet~ provides the best trade-off between the achieved improvements and the decrease of the number of geometric outliers.

\subsection{Comparison with intra-observer variability}
\label{sec:comparison_intra_observer_variability}

As for the segmentation scores, the \lunet~model manages to reach the intra-observer variability for the \LVepi~border ($d_m$ and $d_H$ metrics). The number of geometric outliers, $11$\%, is also reduced below the intra-observer rate. To the best of our knowledge, this is the first time that such result is obtained in the context of 2D echocardiographic image segmentation. \modified{In addition, one can observe that the scores reached by our model are still slightly higher than the intra-observer variability for the \LVendo~border.}

Concerning the estimation of the clinical metrics, although \lunet~improves the results compared to the baseline \unetone~model, its scores are still slightly higher than the intra-observer variability. This reveals that while attention-based networks clearly enhanced the results produced by the baseline \unetone~model, there still exists room for improvement to faithfully reproduce the manual annotations of one expert. 

\subsection{Areas for improvement}
\label{sec:what_sources_of_improvements}

We identified two leads of potential improvement to allow competitive results with respect to the intra-observer variability. First, based on \Tab{localization_results}, it appears that the localization step can be further optimized to improve the \lunet~scores, as suggested by the results on ideal cases provided in \Tab{segmentation_good_medium}. Secondly, there is a need to introduce temporal coherency into deep learning architectures. Indeed, while the current strategy (\ie ED and ES are treated independently) provides high correlations for the estimation of the \LVedv~and \LVesv~($0.956$ for both indices), the estimation of the \LVef is degraded to $0.829$ . This reveals the lack of temporal consistency of the \lunet~segmentation results between ED and ES.

\section{Conclusions}
\label{sec:conclusions}

More accurate and reproducible data analysis is a key innovation in echocardiography, for both diagnosis and patient follow-up. In this study, we introduced a novel multi-task approach to improve the robustness of the segmentation of the endocardium and epicardium in 2D echocardiography. We showed that the joint optimization of the localization and the segmentation tasks leads to better segmentation results at the end of the process. Our method \emph{i)} outperforms \unetone, the current best performing deep learning solution on the CAMUS dataset; \emph{ii)} produced among the best results from the tested attention-based networks; \emph{iii)} produced overall segmentation scores lower than the intra-observer variability for the epicardial border with $11$\% of outliers; \emph{iv)} closely reproduces the expert analysis for the end-diastolic and end-systolic left ventricular volumes, with a mean correlation of $0.96$; \emph{v)} improves the estimation of the ejection fraction of the left ventricle, with scores that remain slightly \modified{higher} than the intra-observer's ones. Though the intra-variability remains to be reached for a set of metrics, this study established localization as a lead for more robust 2D echocardiographic image analysis with a deep learning approach.

\section*{Acknowledgment}

We would like to thank Dr. Ozan Oktay for his help in the implementation of AG-U-Net. This work was performed within the framework of the LABEX PRIMES (ANR- 11-LABX-0063) of Université de Lyon, within the program "Investissements d'Avenir" (ANR-11-IDEX-0007) operated by the French National Research Agency (ANR). The Centre for Innovative Ultrasound Solutions (CIUS) is funded by the Norwegian Research Council (project code 237887).

\bibliographystyle{IEEEtran}
\bibliography{IEEEabrv,lunet_arxiv_2020}

\begin{thebibliography}{10}
\providecommand{\url}[1]{#1}
\csname url@samestyle\endcsname
\providecommand{\newblock}{\relax}
\providecommand{\bibinfo}[2]{#2}
\providecommand{\BIBentrySTDinterwordspacing}{\spaceskip=0pt\relax}
\providecommand{\BIBentryALTinterwordstretchfactor}{4}
\providecommand{\BIBentryALTinterwordspacing}{\spaceskip=\fontdimen2\font plus
\BIBentryALTinterwordstretchfactor\fontdimen3\font minus
  \fontdimen4\font\relax}
\providecommand{\BIBforeignlanguage}[2]{{%
\expandafter\ifx\csname l@#1\endcsname\relax
\typeout{** WARNING: IEEEtran.bst: No hyphenation pattern has been}%
\typeout{** loaded for the language `#1'. Using the pattern for}%
\typeout{** the default language instead.}%
\else
\language=\csname l@#1\endcsname
\fi
#2}}
\providecommand{\BIBdecl}{\relax}
\BIBdecl

\bibitem{Bernard2016}
O.~Bernard, J.~G. Bosch, B.~Heyde, M.~Alessandrini, D.~Barbosa,
  S.~Camarasu-Pop, F.~Cervenansky, S.~Valette, O.~Mirea \emph{et~al.},
  ``{Standardized Evaluation System for Left Ventricular Segmentation
  Algorithms in 3D Echocardiography},'' \emph{IEEE Transactions on Medical
  Imaging}, vol.~35, no.~4, pp. 967--977, 2016.

\bibitem{Pedrosa2017}
J.~Pedrosa, S.~Queirós, O.~Bernard, J.~Engvall, T.~Edvardsen, E.~Nagel, and
  J.~D’hooge, ``{Fast and Fully Automatic Left Ventricular Segmentation and
  Tracking in Echocardiography Using Shape-Based B-Spline Explicit Active
  Surfaces},'' \emph{IEEE Transactions on Medical Imaging}, vol.~36, no.~11,
  pp. 2287--2296, 2017.

\bibitem{Oktay2018}
O.~Oktay, E.~Ferrante, K.~Kamnitsas, M.~Heinrich, W.~Bai, J.~Caballero, S.~A.
  Cook, A.~de~Marvao, T.~Dawes, D.~P. O‘Regan, B.~Kainz, B.~Glocker, and
  D.~Rueckert, ``{Anatomically Constrained Neural Networks (ACNNs): Application
  to Cardiac Image Enhancement and Segmentation},'' \emph{IEEE Transactions on
  Medical Imaging}, vol.~37, no.~2, pp. 384--395, 2018.

\bibitem{Leclerc_tmi_2019}
S.~{Leclerc}, E.~{Smistad}, J.~{Pedrosa}, A.~{Østvik}, F.~{Cervenansky},
  F.~{Espinosa}, T.~{Espeland}, E.~A.~R. {Berg}, P.~{Jodoin}, T.~{Grenier},
  C.~{Lartizien}, J.~{D’hooge}, L.~{Lovstakken}, and O.~{Bernard}, ``Deep
  learning for segmentation using an open large-scale dataset in 2d
  echocardiography,'' \emph{IEEE Transactions on Medical Imaging}, vol.~38,
  no.~9, pp. 2198--2210, Sep. 2019.

\bibitem{Ronneberger2015}
O.~Ronneberger, P.~Fischer, and T.~Brox, ``{U-Net: Convolutional Networks for
  Biomedical Image Segmentation},'' in \emph{Proc. MICCAI}, 2015, pp. 234--241.

\bibitem{zhou2018}
Z.~Zhou, M.~Siddiquee, N.~Tajbakhsh, and J.~Liang, ``Unet++: A nested u-net
  architecture for medical image segmentation,'' in \emph{in proc. of Deep
  Learning in Medical Image Analysis and Multimodal Learning for Clinical
  Decision Support}, 2018, pp. 3--11.

\bibitem{newell2016}
A.~Newell, K.~Yang, and J.~Deng, ``Stacked hourglass networks for human pose
  estimation,'' in \emph{Computer Vision -- ECCV 2016}, B.~Leibe, J.~Matas,
  N.~Sebe, and M.~Welling, Eds.\hskip 1em plus 0.5em minus 0.4em\relax Cham:
  Springer International Publishing, 2016, pp. 483--499.

\bibitem{cvpr_wang_2017}
F.~{Wang}, M.~{Jiang}, C.~{Qian}, S.~{Yang}, C.~{Li}, H.~{Zhang}, X.~{Wang},
  and X.~{Tang}, ``Residual attention network for image classification,'' in
  \emph{2017 IEEE Conference on Computer Vision and Pattern Recognition
  (CVPR)}, 2017, pp. 6450--6458.

\bibitem{nips_ren_fast_r_cnn_2015}
S.~Ren, K.~He, R.~Girshick, and J.~Sun, ``Faster r-cnn: Towards real-time
  object detection with region proposal networks,'' in \emph{Advances in Neural
  Information Processing Systems 28}, 2015, pp. 91--99.

\bibitem{cvpr_redmon_2016}
J.~{Redmon}, S.~{Divvala}, R.~{Girshick}, and A.~{Farhadi}, ``You only look
  once: Unified, real-time object detection,'' in \emph{2016 IEEE Conference on
  Computer Vision and Pattern Recognition (CVPR)}, 2016, pp. 779--788.

\bibitem{iccv_he_2017}
K.~{He}, G.~{Gkioxari}, P.~{Dollár}, and R.~{Girshick}, ``Mask r-cnn,'' in
  \emph{2017 IEEE International Conference on Computer Vision (ICCV)}, 2017,
  pp. 2980--2988.

\bibitem{stacom_payer_2017}
C.~Payer, D.~{\v{S}}tern, H.~Bischof, and M.~Urschler, ``Multi-label whole
  heart segmentation using cnns and anatomical label configurations,'' in
  \emph{Statistical Atlases and Computational Models of the Heart. ACDC and
  MMWHS Challenges}, M.~Pop, M.~Sermesant, P.-M. Jodoin, A.~Lalande, X.~Zhuang,
  G.~Yang, A.~Young, and O.~Bernard, Eds., 2018, pp. 190--198.

\bibitem{prl_guan_2018}
Q.~Guan and Y.~Huang, ``Multi-label chest x-ray image classification via
  category-wise residual attention learning,'' \emph{Pattern Recognition
  Letters}, 2018.

\bibitem{miccai_wang_2018}
Y.~Wang, Z.~Deng, X.~Hu, L.~Zhu, X.~Yang, X.~Xu, P.-A. Heng, and D.~Ni, ``Deep
  attentional features for prostate segmentation in ultrasound,'' in
  \emph{Medical Image Computing and Computer Assisted Intervention -- MICCAI
  2018}, A.~F. Frangi, J.~A. Schnabel, C.~Davatzikos, C.~Alberola-L{\'o}pez,
  and G.~Fichtinger, Eds.\hskip 1em plus 0.5em minus 0.4em\relax Cham: Springer
  International Publishing, 2018, pp. 523--530.

\bibitem{oktay_midl_2018}
O.~Oktay, J.~Schlemper, L.~L. Folgoc, M.~Lee, M.~Heinrich, K.~Misawa, K.~Mori,
  S.~McDonagh, N.~Y. Hammerla, B.~Kainz, B.~Glocker, and D.~Rueckert,
  ``{Attention U-Net: Learning Where to Look for the Pancreas },'' in
  \emph{Medical Imaging with Deep Learning (MIDL'18)}, 2018.

\bibitem{stacom_li_2019}
C.~Li, Q.~Tong, X.~Liao, W.~Si, Y.~Sun, Q.~Wang, and P.-A. Heng, ``Attention
  based hierarchical aggregation network for 3d left atrial segmentation,'' in
  \emph{Statistical Atlases and Computational Models of the Heart. Atrial
  Segmentation and LV Quantification Challenges}, M.~Pop, M.~Sermesant,
  J.~Zhao, S.~Li, K.~McLeod, A.~Young, K.~Rhode, and T.~Mansi, Eds.\hskip 1em
  plus 0.5em minus 0.4em\relax Cham: Springer International Publishing, 2019,
  pp. 255--264.

\bibitem{media_vigneault_2018}
D.~M. Vigneault, W.~Xie, C.~Y. Ho, D.~A. Bluemke, and J.~A. Noble, ``Omega-net:
  Fully automatic, multi-view cardiac mr detection, orientation, and
  segmentation with deep neural networks,'' \emph{Medical Image Analysis},
  vol.~48, pp. 95 -- 106, 2018.

\bibitem{media_pesce_2019}
E.~Pesce, S.~J. Withey, P.-P. Ypsilantis, R.~Bakewell, V.~Goh, and G.~Montana,
  ``Learning to detect chest radiographs containing pulmonary lesions using
  visual attention networks,'' \emph{Medical Image Analysis}, vol.~53, pp. 26
  -- 38, 2019.

\bibitem{media_schlemper_2019}
J.~Schlemper, O.~Oktay, M.~Schaap, M.~Heinrich, B.~Kainz, B.~Glocker, and
  D.~Rueckert, ``Attention gated networks: Learning to leverage salient regions
  in medical images,'' \emph{Medical Image Analysis}, vol.~53, pp. 197 -- 207,
  2019.

\bibitem{nips_jaderberg_2015}
M.~Jaderberg, K.~Simonyan, A.~Zisserman, and k.~kavukcuoglu, ``Spatial
  transformer networks,'' in \emph{Advances in Neural Information Processing
  Systems 28}, 2015, pp. 2017--2025.

\bibitem{leclerc_ius_2019}
S.~{Leclerc}, E.~{Smistad}, J.~{Pedrosa}, A.~{Østvik}, F.~{Cervenansky},
  F.~{Espinosa}, T.~{Espeland}, E.~A.~R. {Berg}, P.~{Jodoin}, T.~{Grenier},
  C.~{Lartizien}, J.~{D’hooge}, L.~{Lovstakken}, and O.~{Bernard}, ``{RU-Net:
  A refining segmentation network for 2D echocardiography},'' in \emph{IEEE
  International Ultrasonics Symposium (IUS)}, 2019.

\bibitem{Folland1979}
E.~D. Folland, A.~F. Parisi, P.~F. Moynihan, D.~R. Jones, C.~L. Feldman, and
  D.~E. Tow, ``{Assessment of left ventricular ejection fraction and volumes by
  real-time, two-dimensional echocardiography. A comparison of cineangiographic
  and radionuclide techniques},'' \emph{Circulation}, vol.~60, no.~4, pp.
  760--766, 1979.

\bibitem{Krizhevsky_nips_2012}
A.~Krizhevsky, I.~Sutskever, and G.~E. Hinton, ``Imagenet classification with
  deep convolutional neural networks,'' in \emph{Advances in Neural Information
  Processing Systems 25}, F.~Pereira, C.~J.~C. Burges, L.~Bottou, and K.~Q.
  Weinberger, Eds., 2012, pp. 1097--1105.

\bibitem{Simonyan2014}
K.~Simonyan and A.~Zisserman, ``Very deep convolutional networks for
  large-scale image recognition,'' \emph{CoRR}, vol. abs/1409.1556, 2014.

\end{thebibliography}

\end{document}